\documentclass[aps,prl,showpacs,twocolumn,groupedaddress]{revtex4}

\usepackage{graphicx}
\usepackage{amssymb}
\usepackage{amsmath}

\newcommand{\rmd}{{\rm d}}

\newcommand{\rmi}{{\rm i}}

\newcommand{\eps}{\varepsilon}

\newcommand{\la}{\langle}
\newcommand{\ra}{\rangle}

\begin{document}

\title{Anomalous slow fidelity decay for symmetry breaking perturbations}

\author{T.~Gorin$^1$, H.~Kohler$^2$, T.~Prosen$^3$, T.~H.~Seligman$^4$,
H.-J.~St\" ockmann$^5$ and M.~\v Znidari\v c$^{3,6}$}

\affiliation{
$^1$ Max-Planck Institute f\" ur Physik der Komplexer Systeme, D-01187 Dresden, 
     Germany\\
$^2$ Institut f\" ur theoretische Physik, Universit\"at Heidelberg, D-69120 
     Heidelberg, Germany\\
$^3$ Department of Physics, Faculty of Mathematics and Physics, University of 
     Ljubljana, SI-1000 Ljubljana, Slovenia\\
$^4$ Centro de Ciencias F\'\i sicas, UNAM, Cuernavaca, Mexico\\
$^5$ Fachbereich Physik der Philipps-Universit\"at Marburg, D-35032 Marburg, 
     Germany\\
$^6$ Abteilung f\" ur Quantenphysik, Universit\" at Ulm, D-89069 Ulm, Germany} 

\date{\today}

\begin{abstract}
Symmetries as well as other special conditions can cause 
  anomalous slowing down of 
  fidelity decay. These situations will be characterized,
  and a family of random matrix models to emulate them generically
  presented. An analytic solution based on exponentiated linear 
  response will be given. For one representative case the exact
  solution is obtained from a supersymmetric calculation. The 
  results agree well with dynamical calculations for a kicked top.
\end{abstract}


\maketitle

Sensitivity to perturbations as measured by fidelity decay has received
a great deal of attention both in the context of quantum and classical
dynamical systems~\cite{general,Pro02,ProZni02,PSZ03} and as a benchmark 
for the stability of
possible quantum information devices~\cite{NieChu00,ProZni01}.
Fidelity
may be described as the cross-correlation function between unperturbed and 
perturbed evolution of a quantum or classical wave system.
Linear response
theory has been particularly successful in describing fidelity decay,
by relating it to the correlation decay of the perturbing operator
in the interaction picture~\cite{ProZni01,Pro02,ProZni02}.
For chaotic systems a recently introduced random matrix model~\cite{GPS04} 
is in excellent agreement with experiments~\cite{SSGS05,GSW05}.

It has been shown, that so-called residual perturbations  that
have vanishing diagonal matrix elements in the eigenbasis of the
unperturbed Hamiltonian, lead to very slow fidelity
decay known also as quantum freeze of fidelity~\cite{PZ03,PZ05}.
Fidelity freeze is a pure wave phenomenon without classical 
analogue~\cite{PZ05}. 
In the field of quantum computation, it is known as {\em dynamical 
decoupling}~\cite{viola,review}.
One can identify four different physical situations in which quantum 
freeze or anomalous slow fidelity decay occurs. The case when the
perturbation can be written as a time derivative (or commutator with 
the unperturbed Hamiltonian) is treated in Refs.~\cite{PZ03,PZ05}.

In this letter we treat the three other and indeed physically most important 
cases: The first one corresponds to a perturbation which breaks 
an antiunitary symmetry (e.g. time-reversal) in an optimal way, meaning that 
the perturbation anticommutes with the antiunitary symmetry, e.g.
switching on the magnetic field. The other two cases correspond to a 
{\em mean field approach} in which the diagonal part of the perturbation is 
moved to the unperturbed Hamiltonian. 
We will consider unperturbed Hamiltonians, with and without an
antiunitary symmetry. To obtain a generic understanding, all cases are 
considered in the framework of random matrix theory (RMT).
We present general theoretical results in the linear respons regime, as well as
an exact analytical result obtained by a supersymmetry method~\cite{sk} for 
the antiunitary symmetry-breaking case.
Our results display excellent agreement with
numerical experiments for quantum kicked tops.

To study fidelity decay, we consider the perturbed Hamiltonian 
$H=H_0 + \lambda V$. If $U(t)$ and $U_0(t)$ are the unitary propagators
under $H$ and $H_0$, respectively, we define the fidelity amplitude as
\begin{equation}
f(t) = \la\Psi_0(t)|\Psi(t)\ra  = \la\Psi(0)| U_0(-t)\,
U(t)|\Psi(0)\ra \; , 
\label{defH}\end{equation}
where $\Psi(t)= U(t)\, \Psi(0)$ and $\Psi_0(t)= U_0(t)\, \Psi(0)$.
Fidelity is defined as $F(t)= |f(t)|^2$. 
In Refs.~\cite{Pro02,ProZni02}, it has been shown that within second order 
time-dependent perturbation theory (Born series) the fidelity
amplitude, averaged over random initial states, can
be expressed in terms of the two-point time correlation integral
${\cal C}(t)$:
\begin{equation}\label{linres}
\la f(t)\ra_E = 1 - 4\pi^2\lambda^2\; {\cal C}(t) + {\cal O}(\lambda^4) \; ,
\end{equation}
where $\la\ldots\ra_E$ denotes the average over random initial states, which
are concentrated on a small energy interval that contains many levels $N_E$.
For the fidelity
amplitude, this averaging amounts to taking a restricted trace of the echo 
operator, in the eigenbasis of the unperturbed Hamiltonian $H_0$. Therefore, 
we may write:
\begin{align}
{\cal C}(t) &= \int_0^t \rmd t' \int_0^{t'} \rmd t'' \la V(t')V(t'')\ra_E \\
 \la V(t')V(t'')\ra_E &= \frac{1}{N_E}{\sum_\alpha}'\sum_\beta 
   |V_{\alpha\beta}|^2\, e^{2\pi\rmi (E_\alpha - E_\beta)(t'-t'')},
\notag\end{align}
where the $E_\alpha$ are the eigenvalues of $H_0$, 
and $V(t) = U_0(t)^\dagger\, V\, U_0(t)$ is the perturbation in the interaction
picture. The matrix elements  $V_{\alpha\beta}$ are taken in the eigenbasis
of $H_0$. The primed sum 
runs over $N_E$ eigenstates of $H_0$.

Often, the exponentiated version of Eq.~(\ref{linres}),
\begin{equation}\label{linresexp}
\la f(t)\ra_E = e^{- \eps{\cal C}(t)} + {\cal O}(\lambda^4)\, , \qquad
\eps=4\pi^2\lambda^2\, ,
\end{equation}
is able to describe the fidelity decay from the perturbative to
the golden rule regime well~\cite{GPS04,SSGS05,GSW05,Hau05}. This
is true, in particular, if the system is chaotic, such that the perturbation 
can be described~\cite{GPS04} by one of the random Gaussian ensembles (RMT 
approach)~\cite{Meh91}.

We now consider the situation of a residual perturbation, i.\,e.,
one with vanishing diagonal, $V_{\alpha \alpha} \equiv 0$ within the RMT
models. Thus, we assume that the non-zero matrix elements of this perturbation 
are independent normalized Gaussian random variables with variance
\begin{equation}
\la |V_{\alpha \beta}|^2 \ra = 1-\delta_{\alpha \beta} \; ,
\label{V:statdef}\end{equation}
where $\la\ldots\ra$ denotes an ensemble average.
For the perturbation matrix $V$, we consider three different ensembles, which 
fulfill Eq.~(\ref{V:statdef}): 
(i) an ensemble of imaginary antisymmetric matrices,
(ii) an ensemble of real symmetric matrices with deleted diagonal, 
and (iii) an ensemble of Hermitean matrices with deleted diagonal.

The average of the correlation integral $\mathcal{C}(t)$ over any of those
ensembles gives:
\begin{equation}
\la {\cal C}(t)\ra = \frac{t}{2} - \int_0^t \rmd t'\int_0^{t'} \rmd t''
b(t'') \label{LR:corrint}
\end{equation}
where $b(t)$ is the two-point spectral form factor of $H_0$. Here, as well as
throughout the rest of the letter, we use units, where the Heisenberg time is
equal to one. If $N_E$ is sufficiently large,
$b(t)$ tends to a well defined smooth function (self-averaging); 
else we average over a random matrix ensemble for $H_0$ as well. 
For Gaussian orthogonal (GOE) and unitary (GUE) ensembles, the form factors 
are given in Ref.~\cite{Meh91}.  Note that the term
proportional to $t^2$ is missing as compared to the linear
response result for a generic perturbation~\cite{GPS04}. 
For what follows, we assume that the average $\la\ldots\ra$ includes such a 
procedure.



Let us first concentrate on the special case of unperturbed
Hamiltonians $H_0$ taken from the GUE. Then, 
we have $b(t) = {\rm max}\{1-t,0\}$, and
\begin{equation}
\la\mathcal{C}(t)\ra = {\cal C}_{\rm GUE}(t) = \begin{cases}
\frac{t}{2}-\frac{t^2}{2}+\frac{t^3}{6} \; &: t \le 1, \\
 \frac{1}{6} \; &: t > 1 \end{cases}
\label{eq:cGUE}\end{equation}
As a result, for times longer than the Heisenberg time, Eq.~(\ref{linres}) 
predicts the fidelity to ``freeze'' on a minimal value
\begin{equation}
f_{\rm plateau} = 1 - \frac{\eps}{6} \; .
\label{GUEfreezelev}\end{equation}
Since the next correction term grows quadratically in time, 
$\la f(t)\ra = f_{\rm plateau} + {\cal O}(\lambda^4 t^2)$, we find that the
plateau ends at a time of order $t^* \sim 1/\lambda$.

Second, we choose $H_0$ from the GOE. Also in this case, the
integrals in equation~(\ref{LR:corrint}) can be performed
analytically (see Ref.~\cite{GPS04}). Here we just give the the
leading asymptotics for $t \gg 1$:
\begin{equation}
{\cal C}_{\rm GOE}(t) = \frac{\ln(2t) + 2}{12} + {\cal O}(t^{-1}\, \ln t)
\label{eq:cGOE}
\end{equation}
This yields a logarithmically slow decay of fidelity
\begin{equation}
\la f(t) \ra \approx 1 - \frac{\eps}{12}\left[2 + \ln(2t)\right] + {\cal
O}(\lambda^4 t^2) \; .
\label{GOEfreezelev}\end{equation}
Both results for the plateau value of the fidelity amplitude,
Eq.~(\ref{GUEfreezelev}) and Eq.~(\ref{GOEfreezelev}),  follow from 
Eq.~(\ref{LR:corrint}). They are thus valid for
any of the three ensembles used for the perturbation, (i), (ii), and (iii).
Indeed, a similar result could be obtained for the Gaussian symplectic 
ensemble.

In~\cite{GPS04} the averages $\la F(t)\ra$ and $|\la f(t)\ra|^2$ were shown 
to differ only in a term proportional to $t^2$, which is absent in the quantum
freeze case. Thus, the eqs.~(\ref{linres}) and (\ref{LR:corrint}) yield the 
linear response approximation for fidelity decay. 
Numerics indicate that $\la F(t)\ra \approx |\la f(t)\ra|^2$, in general, in 
accordance with an
argument given in~\cite{PSZ03}.

For long times and small perturbations, the fidelity
amplitude can be expressed in terms of level shifts. 
Within the second order stationary perturbation theory, its average is then
given by the Fourier transform of the level curvature
distribution~\cite{review} which was obtained analytically
in~\cite{FyoSom95}. The final result is surprisingly simple:
\begin{equation}
\la f(t)\ra = \begin{cases} \tau\; K_1(\tau) &: \text{GOE}\\
   (1+\tau)\; {\rm e}^{- \tau} &: \text{GUE}\end{cases}\qquad
\tau = \frac{\eps t}{2} \; .
\label{eq:asymp}
\end{equation}
where $K_1$ is the modified Bessel function of first order. 
The GOE branch is valid for an unperturbed GOE Hamiltonian, and a 
perturbation matrix of type (i) or (ii). The GUE branch is valid for an
unperturbed GUE Hamiltonian, and a perturbation matrix of type (iii).
More details on the asymptotic behavior of fidelity decay in situations
of quantum freeze will be published elsewhere~\cite{review}.

One should stress that diagonal elements of the perturbation vanish also
in the presence of a discrete or continuous {\em unitary} symmetry $R$, 
of $H_0$, which anti-commutes with $V$, $RV=-VR$. 
However, it turns out that its effect on fidelity enhancement is less 
drastic than the predictions of Eqs.~(\ref{eq:cGUE}) and~(\ref{eq:cGOE}), 
because of the lack of
correlations between different subspectra of $H_0$. As a result,
the asymptotic growth of the correlation integral is linear ${\cal
C}(t) \propto t$, for times before and after the Heisenberg time.

For the case of $H_0$ taken from the GOE and a purely imaginary
antisymmetric perturbation the average fidelity
can be obtained exactly by supersymmetry techniques in the
limit of large dimension $N$.
The calculation is technically much more involved than for the
case of a GOE perturbation~\cite{stoe04b} and will be presented
elsewhere~\cite{sk}. The result again is a VWZ-like integral (see
Ref.~\cite{VWZ}, Eq. (8.10)) and is given by
\begin{align}\label{09}
  \la f(t)\ra &=2\!\!\!\!\!\!\!
  \int\limits_{{\rm Max}(0,t-1)}^t\!\!\!\!\!\!\!\! du\int\limits_0^u
  \frac{v\,dv}{\sqrt{[u^2-v^2][(u+1)^2-v^2]}} \notag\\
&\quad\times\frac{(t-u)(1-t+u)}{(v^2-t^2)^2}[1+\eps(t^2-v^2)]
\notag\\
&\quad\times [t(2u+1-t)+v^2]
e^{-\frac{\eps}{2}[t(2u+1-t)-v^2]}\,.
\end{align}
%
The only difference to Ref.~\cite{stoe04b}, where a GOE
perturbation was considered, is the additional factor
$[1+\eps(t^2-v^2)]$ in the integrand, and a minus sign with
the $v^2$ term in the exponent, where in the GOE case there is a
plus sign. 
%
%

\begin{figure}
\includegraphics[width=\linewidth]{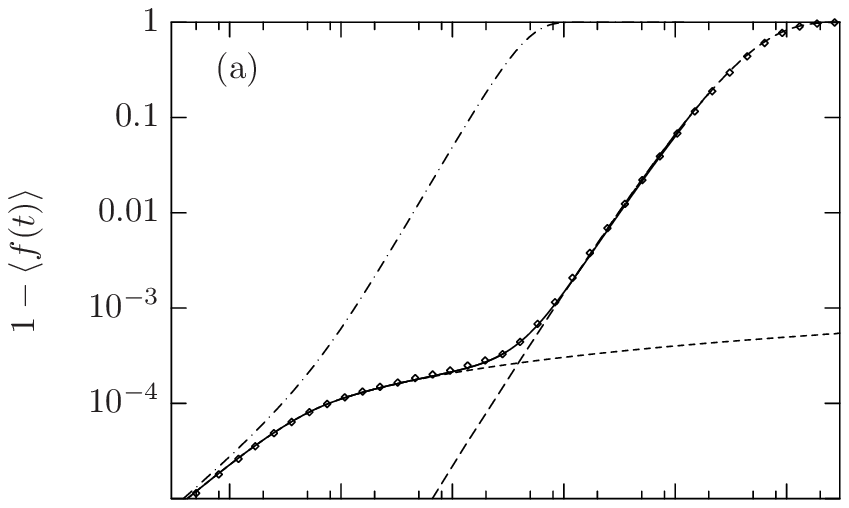}

\vspace{-2ex}

\includegraphics[width=\linewidth]{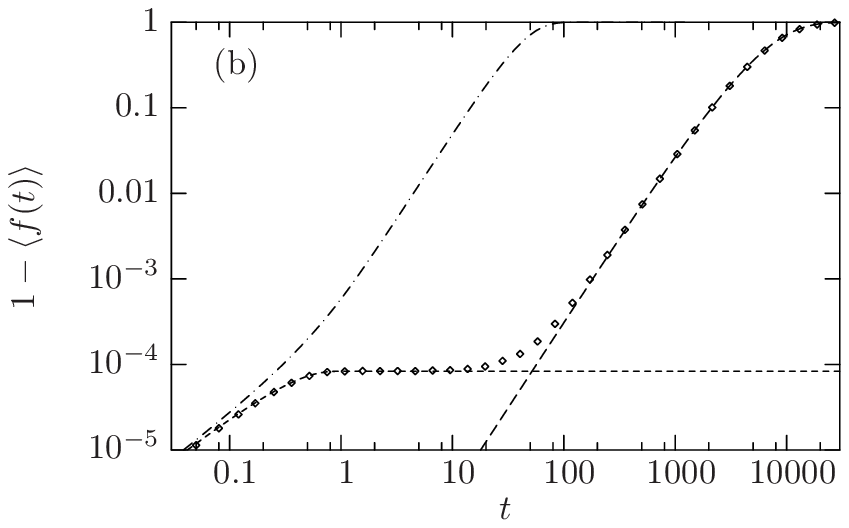}
\caption{The complement of the average fidelity amplitude,
for a weak perturbation, $\eps= 5\times 10^{-5}$. 
Part (a) shows the GOE case with a purely
antisymmetric random Gaussian perturbation. Part (b) shows the GUE case, with
an independent GUE perturbation with deleted diagonal. The exponentiated
linear response approximations are plotted with short dashed lines (for 
comparison, the exact results for full GOE and GUE perturbations are shown
with chain curves).
The results from time-independent perturbation theory are plotted with long
dashed lines. In the GOE case, the exact analytical result, Eq.~(\ref{09}) is
plotted with a solid line. The random matrix simulations are plotted with
points.}
\label{fig:freeze}\end{figure}

Figure \ref{fig:freeze} shows the fidelity decay both for $H_0$
taken from the GOE (a) and the GUE (b), for a small perturbation 
$\eps= 5\times 10^{-5}$. It compares different theoretical approaches.
To this end, one minus the average
fidelity amplitude is plotted on a double-log scale.
The figure shows the exact result 
calculated from Eq.~(\ref{09}) (in the GOE case only),
together with the results from the exponentiated linear response 
approximation (\ref{linresexp}) and
the asymptotic expression (\ref{eq:asymp}). For comparison the
fidelity decay for the case of a GOE perturbation~\cite{stoe04b}
is shown as well. We see that the linear response approximation is
able to describe the fidelity decay for quite a long time very
well. 
Immediately beyond the time when the linear response formulae fail, the
asymptotic results (\ref{eq:asymp}) describe fidelity decay quite well.
We also compare to numerical random matrix 
simulations, where we computed averages over 
$10^4$ samples of $100\times100$ matrices. Only
the 10 central states have been taken into account. 
In the GOE case, we have also performed numerical simulations for symmetric
perturbations with deleted diagonal (case (ii)) and we have not found
significant deviations from the antisymmetric case (i). 
For strong perturbations the plateau disappears and we find a partial revival
of fidelity near the Heisenberg time, similar as in Ref.~\cite{stoe04b}.

We have concentrated on the unitary antisymmetric perturbation,
because of its invariance properties, which allow to obtain
results in closed form. Yet the linear response result in Eq.~(\ref{linres}) 
can be carried to higher order, and at
least up to sixth order they coincide with the one for symmetric
perturbations with deleted diagonal~\cite{review}. 

The RMT model can also be compared to dynamical
systems with chaotic classical limit. For this purpose we have
considered a quantized kicked top~\cite{Haake:87}.

In the first example, we choose a one step propagator
\begin{equation}
U_\lambda=P^{\frac{1}{2}} {\rm e}^{-\rmi \gamma S_{\rm
y}}P^{\frac{1}{2}} {\rm e}^{-\rmi \lambda S_{\rm x}},
\quad
\gamma = \pi/2.4
\label{eq:U0anti}
\end{equation}
with $P={\rm e}^{-\rmi \alpha S_{\rm z}^2/2S} {\rm e}^{-\rmi
S_{\rm z}}$ and $S_{\rm x,y,z}$ being standard spin operators. 
$U_0$ is time-reversal invariant, and the perturbation $S_{\rm y}$
is antisymmetric in the eigenbasis of $U_0$.
The ``symmetrization'' of $U_0$ is essential for $V$ to anticommute with the 
time-reversal symmetry.

\begin{figure}
\includegraphics[angle=-90,width=\linewidth]{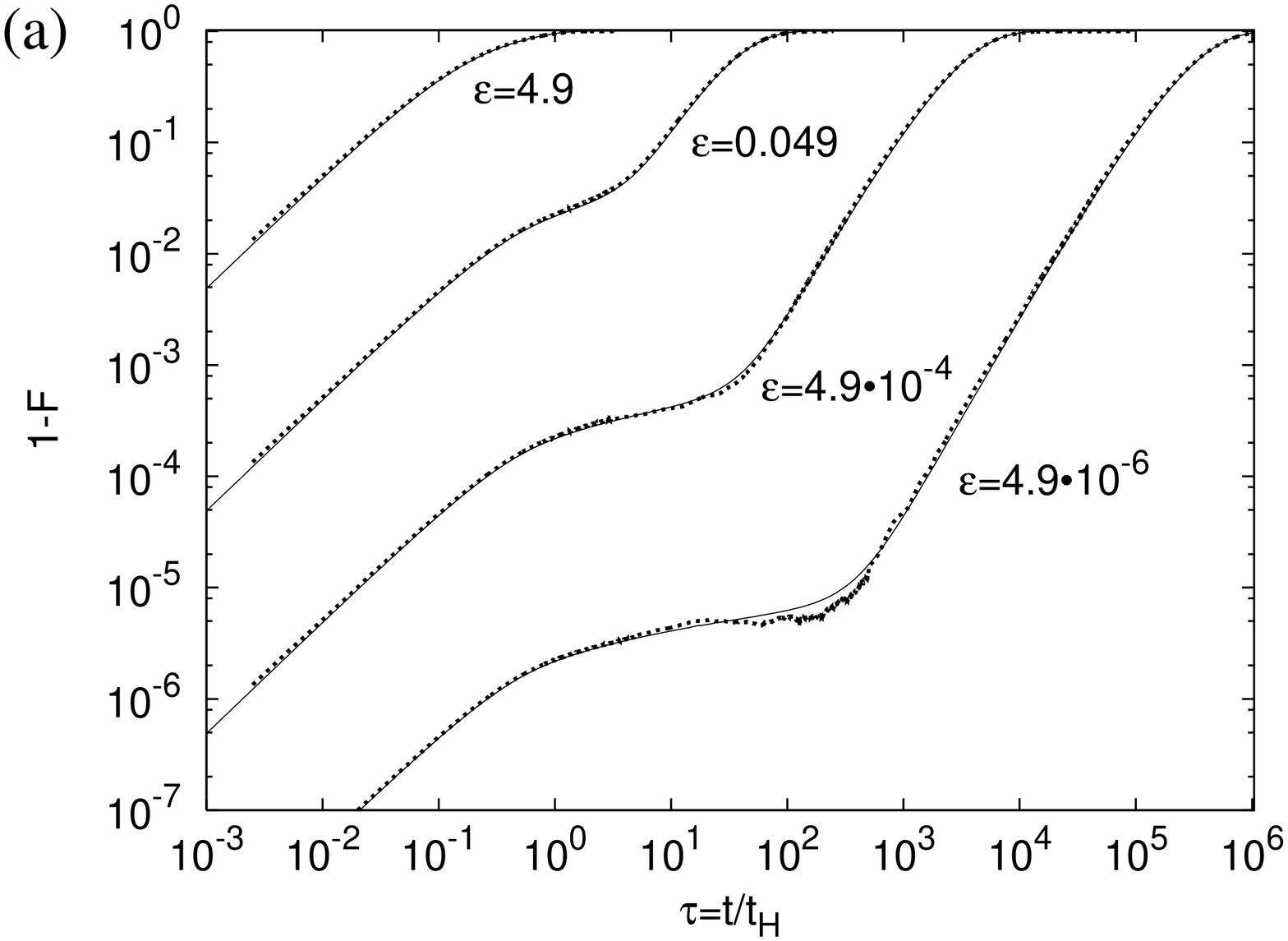}\\
\includegraphics[angle=-90,width=\linewidth]{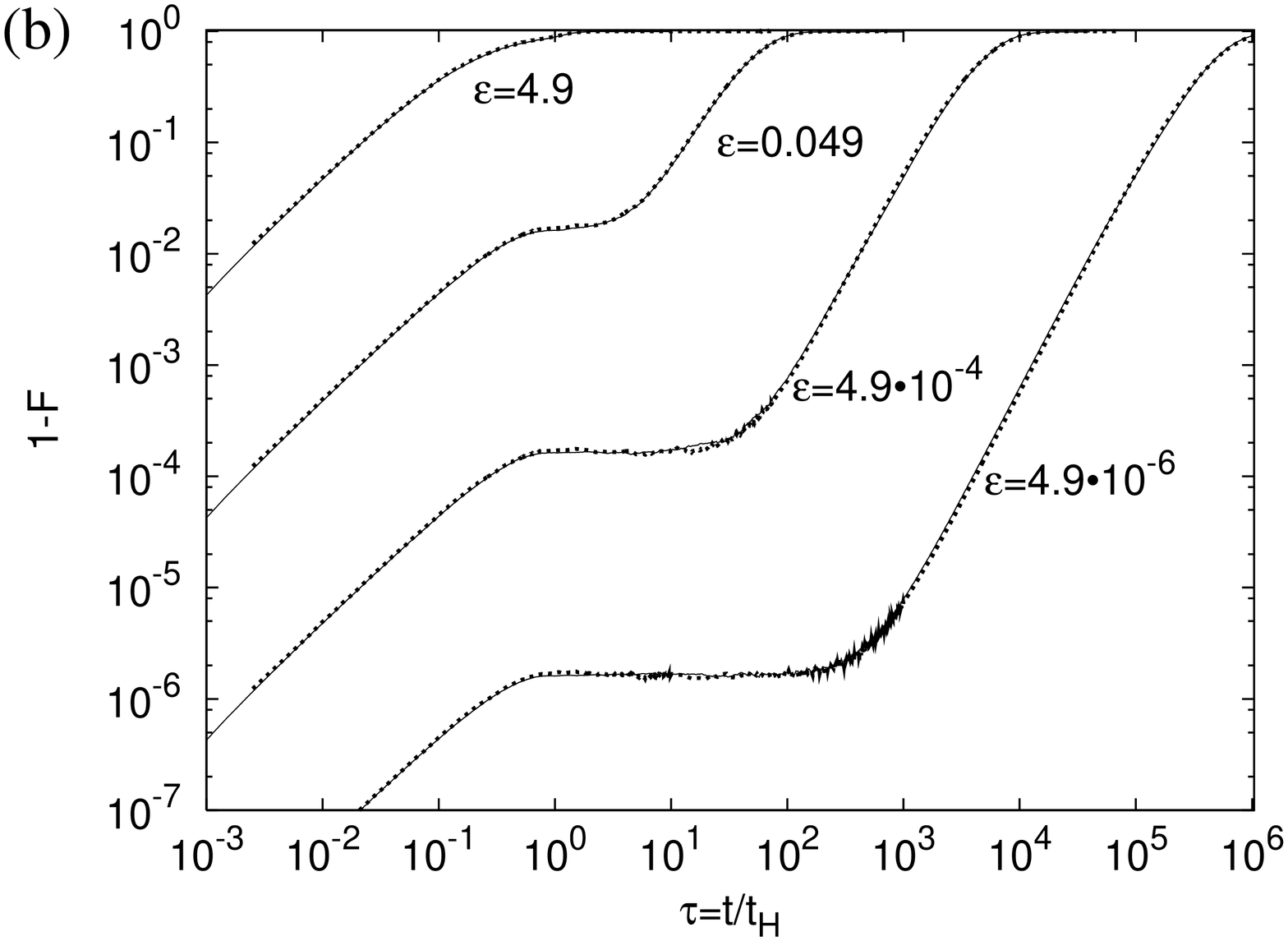}
\caption{Fidelity freeze for a quantized kicked top. In (a) perturbation is 
imaginary antisymmetric with unperturbed dynamics having antiunitary symmetry 
(\ref{eq:U0anti}) while in (b) the unperturbed evolution as well as the 
perturbation have 
no symmetries left (\ref{eq:guektop}). The dashed lines give the 
numerical simulations, while the solid lines give the square of the 
theoretical prediction (\ref{09}) in (a) and RMT simulations in (b).
}
\label{fig:ktop}
\end{figure}

We choose the spin $S=200$, one initial random state and average the fidelity
over $400$ realizations of the propagator $U_\lambda$ where for each
realization we draw a parameter $\alpha$ from a Gaussian distribution of width 1 
centered around 30. The results of fidelity decay for different
strengths of perturbation are shown in Fig.~\ref{fig:ktop}(a). 
We find good agreement with the square of the theoretical result (\ref{09})
for the fidelity amplitude, which in turn agrees well with RMT simulations for
the fidelity (not shown).
Without averaging over an
ensemble of dynamical systems we get considerable fluctuations around
RMT curves. Note that we do not use any fit parameters. 
The dimensionless perturbation strength $\eps$ in
Eq.~(\ref{09}) is obtained as $\eps=2N\sigma_{\rm cl}(S
\lambda)^2=4\lambda^2 S^3 \sigma_{\rm cl}$, where $\sigma_{\rm
cl}=0.153$ is an integral of the classical correlation function
calculated using the corresponding classical map, see {\em e.g.}
Ref.~\cite{ProZni02} for more details. Heisenberg time is $t_{\rm
H}=N=2S$. 

We also consider an unperturbed propagator without 
time-reversal symmetry, that corresponds to GUE case,
\begin{equation}
U_\lambda=P {\rm e}^{-\rmi \gamma S_{\rm y}} \, 
   {\rm e}^{-\rmi \mu S_{\rm x}^2/2S} \, {\rm e}^{-\rmi \xi S_{\rm x}} \,
{\rm e}^{-\rmi \lambda S'_{\rm x}} \, .
\label{eq:guektop}\end{equation}
with $\gamma=\pi/2.4,\mu=10,\xi=1$.
Here we have set diagonal matrix elements of the perturbation in
the eigenbasis of $U_0$ to zero by hand, 
$S'_{\rm x} = S_{\rm x} - {\rm diag}S_{\rm x}$.
We take $S=200$ and average the fidelity over $100$ samples, 
similarly as for GOE case. 
As above we determine $\eps$ from the classical correlation integral
$\sigma_{\rm cl}=0.168$. In 
Fig.~\ref{fig:ktop}b the results of the numerical simulation are plotted, 
together with RMT Monte Carlo simulation (full line). Again, good agreement 
with the RMT model is observed. 

We have presented RMT models that display the eminent features
of quantum freeze of fidelity under a wide range of circumstances not 
previously considered. We allow for any unperturbed Hamiltonian or ensemble
of Hamiltonians for which the spectral form factor is known.
The perturbations are represented by ensembles of random Hermitean
matrices with zero entries on the diagonal.
We give a perturbative solution for the general model, and
we present an exact solution obtained by supersymmetric techniques,
for the case of Hermitean antisymmetric
perturbations of GOE Hamiltonians.
Kicked top models display excellent
agreement with the random matrix results.

The physical importance of such systems becomes apparent in two
quite different aspects. On one hand mean field theories in some
sense include the diagonal part of the perturbation in the
unperturbed Hamiltonian, and thus the quantum freeze sheds new
light on the surprising success of such theories. On the other
hand this result shows that  for a quantum
information process to be effective beyond the Heisenberg time,
one has to suppress the diagonal part of any static perturbation. 

\begin{acknowledgments}
We thank H. Schomerus for drawing our attention to Ref.~\cite{FyoSom95}.
THS acknowledges support by the grants No.~10803 (DGAPA) and No.~41000-F
(CONACyT). MZ thanks the Alexander von Humboldt Foundation. TP acknowledges 
support by the grant P1-0044 of the Slovenian Research Agency.
\end{acknowledgments}

\end{document}